\documentclass[aip,amsmath,amssymb,longbibliography,nofootinbib,reprint, groupedaddress,
]{revtex4-1}

\usepackage{graphicx}
\usepackage{dcolumn}
\usepackage{bm}
\usepackage[breaklinks=true,colorlinks=true,linkcolor=blue,urlcolor=blue,citecolor=blue]{hyperref}

\usepackage[utf8]{inputenc}
\usepackage[T1]{fontenc}
\usepackage{mathptmx}
\usepackage{etoolbox}
\DeclareMathSymbol{\mhyphen}{\mathord}{AMSa}{"39}

\def\ket#1{{\left| #1 \right\rangle}}

\def\c#1{\textrm{#1}}

\makeatletter
\def\@email#1#2{%
 \endgroup
 \patchcmd{\titleblock@produce}
  {\frontmatter@RRAPformat}
  {\frontmatter@RRAPformat{\produce@RRAP{*#1\href{mailto:#2}{#2}}}\frontmatter@RRAPformat}
  {}{}
}%
\makeatother
\begin{document}

\preprint{}

\title[A simple asymptotically optimal Clifford circuit compilation algorithm]{A simple asymptotically optimal Clifford circuit compilation algorithm}

\author{Timothy Proctor}
 \email{tjproct@sandia.gov.}
\author{Kevin Young}
\affiliation{Quantum Performance Laboratory, Sandia National Laboratories, Livermore, CA 94550, USA}

\begin{abstract}
We present an algorithm that decomposes any $n$-qubit Clifford operator into a circuit consisting of three subcircuits containing only \textsc{cnot} or \textsc{cphase} gates with layers of one-qubit gates before and after each of these subcircuits. As with other asymptotically optimal Clifford compilation algorithms, the resulting circuit contains $O(n^2/\log n)$ two-qubit gates. The derivation of our algorithm only requires the symplectic representation of Clifford gates, basic row and column matrix manipulations, and some known properties of general matrices over 0 and 1.
\end{abstract}

\maketitle

\section{Introduction} 
Clifford circuits are important for a variety of quantum information tasks, so algorithms for optimizing circuits containing only Clifford gates are a useful component in the broader toolkit for quantum circuit compilation. It is well-known that an arbitrary $n$-qubit Clifford operator can be compiled into a circuit containing $O(n^2)$ one- and two-qubit gates  \cite{dehaene2003clifford,hostens2005stabilizer}, because this compilation problem can be expressed as Gaussian elimination on $n \times n$ matrices over $\{0,1\}$, where the row operations of the elimination correspond to the application of $\textsc{cnot}$, Hadamard and phase gates. However the $O(n^2)$ scaling of this algorithm is not asymptotically optimal in two-qubit gate count---where, here and throughout, all \textsc{cnot} gate counts use an all-to-all connectivity architecture. First Patel \emph{et al.} \cite{patel2003efficient} showed that $O(n^2/\log n)$ \textsc{cnot} gates are necessary and sufficient to implement an arbitrary \textsc{cnot} circuit---i.e., a circuit that contains only \textsc{cnot} gates---and they provided an explicit algorithm with this scaling. Then Aaronson and Gottesman \cite{aaronson2004improved} showed that an arbitrary Clifford operator can be decomposed into seven \textsc{cnot} subcircuits with a layer of one-qubit gates between each of these subcircuits. This then implies that $O(n^2/\log n)$ one- and two-qubit gates are necessary and sufficient to implement any $n$-qubit Clifford operator. 

Aaronson and Gottesman's compilation algorithm has asymptotically optimal scaling in the number of \textsc{cnot} gates \cite{aaronson2004improved}, but it contains more \textsc{cnot} subcircuits than necessary \cite{maslov2017optimized, Amy2022-gt, Bravyi2020-hg, Bravyi2021-my}. Maslov and Roetteler \cite{maslov2017optimized} presented an algorithm that generates an arbitrary Clifford operator from a circuit containing only four \textsc{cnot} subcircuits. Then Bravyi and Maslov \cite{Bravyi2020-hg} showed how to decompose a Clifford operator into a circuit containing only three subcircuits of two-qubit gates, and Bravyi \emph{et al.} \cite{Bravyi2021-my} presented heuristic algorithms for Clifford compilations that are near-optimal in two-qubit gate count.

In this paper we present an efficient algorithm that decomposes any $n$-qubit Clifford operator into a circuit containing three \textsc{cnot} circuits with layers of one-qubit gates in between. We then present an algorithm for stabilizer state preparation. This algorithm takes an $n$-qubit stabilizer state $\ket{\psi}$ and creates a circuit, that produces $\ket{\psi}$ from $\ket{0}^n$, consisting of a single $\textsc{cnot}$ circuit preceded and followed by single-qubit gates. To the best of our knowledge, our Clifford compilation algorithm does not create circuits with better properties (e.g., fewer two-qubit gates) than that of Bravyi and Maslov \cite{Bravyi2020-hg}, but its derivation is simple: our derivation only uses the symplectic representation of Clifford gates, basic row and column matrix manipulations, and some known properties of general matrices over 0 and 1. Our algorithm is implemented in \texttt{pyGSTi}  \cite{nielsen2020probing}, and it was originally developed for the purpose of making direct randomized benchmarking (DRB) \cite{proctor2018direct} and Clifford randomized benchmarking (CRB) \cite{magesan2011scalable} feasible on more qubits. In particular, the reference implementation of DRB in \texttt{pyGSTi} \cite{nielsen2020probing}, and the first experimental demonstrations of DRB \cite{proctor2018direct}, use the compilation algorithm for stabilizer state preparation circuits presented herein.

\section{Background}\label{sec:background}
\subsection{The symplectic representation of Clifford operators}
Our compilation algorithm uses the symplectic representation of Clifford operators, and so we begin by reviewing this representation \cite{dehaene2003clifford, hostens2005stabilizer}. The $n$-qubit Pauli group consists of all $n$-fold tensor products of the four Pauli operators $I$, $X$, $Y$ and $Z$ with a phase of $\pm i$ or $\pm 1$. As $Y = iZX$, any $n$-qubit Pauli operator can be expressed as:
\begin{equation}
P_{\xi, \vec{v}} = i^{\xi} X^{v_1}Z^{v_{1+n}} \otimes  X^{v_2}Z^{v_{2+n}}  \otimes \cdots \otimes  X^{v_{n}}Z^{v_{2n}}  .
\end{equation}
where $\xi \in \{0,1,2,3\}$ and $\vec{v} \in \{0,1\}^{2n}$. In this notation, the Pauli group is defined by
\begin{equation}
\mathbb{P} = \{P_{\xi,\vec{v}} \mid \xi \in \{0,1,2,3\}, \, \vec{v} \in \{0,1\}^{2n}\}.
\end{equation}
The $n$-qubit Clifford group ($\mathbb{C}$) consists of all unitaries that map Pauli operators to Pauli operators under conjugation:
\begin{equation}
\mathbb{C} = \{ U \in \mathbb{U}(2^n) \mid U P U^{\dagger} \in \mathbb{P} \,\, \forall P \in \mathbb{P} \},
\end{equation}
where $\mathbb{U}(2^n$) denotes the $n$-qubit unitary group.

The elements of the Clifford group can be represented by their action on $\xi$ and $\vec{v}$, which index the Pauli group elements. In particular, a Clifford group element $U$ can be represented by a unique matrix $S(U)$ over the 2-element finite field $\mathbb{F}_2$ (the set $\{0,1\}$ equipped with modulo-2 arithmetic) and a unique $2n$ dimensional vector $p(U)$ over $\{0,1,2,3\}$. This is because $U$ is entirely described by its conjugation action on a minimal set of generators for the Pauli group \cite{hostens2005stabilizer}. So if we define
$\vec{v}_i$ and $\xi_i$ by 
\begin{equation}
U P_{0, \vec{e}_i}  U^{\dagger} =  P_{\xi_i, \vec{v}_i},
\end{equation}
where $\vec{e}_i$ is the $i^{\rm th}$ standard basis vector, i.e., $\vec{e}_0 = (1,0,0,\dots)$ etc, then $U$ can be described by arranging the $\vec{v}_i$ into a matrix and the $\xi_i$ into a vector:
\begin{equation}
S(U) = (\vec{v}_0, \vec{v}_{1}, \dots), \hspace{0.5cm} p(U) = (\xi_0, \xi_1, \dots).
\end{equation}

The Clifford operators are unitaries, and so they must preserve the commutation relations of the Pauli operators. This then implies that $S(U)$ must be symplectic for every Clifford operator $U$. A $2n \times 2n$ matrix $S$ over a field $\mathbb{F}$ is symplectic if
\begin{equation}
S^T \Omega S = \Omega, \label{eq:symp}
\end{equation}
where, here and throughout, the arithmetic is that of $\mathbb{F}$, and $\Omega$ is the symplectic form:
\begin{equation}
\Omega = \begin{pmatrix} 0 & -I_n \\ I_n & 0 \end{pmatrix},
\end{equation}
where $I_n$ is the identity matrix with dimension $d$ (in the remainder of this paper we will typically drop the dimensional subscript on the identity matrix).

Every $2n \times 2n$ symplectic matrix $S$ together with some $p \in  \{0,1,2,3\}^{2n}$ defines a Clifford operator \cite{hostens2005stabilizer}\footnote{For each symplectic matrix $S$, not every $p \in \{0,1,2,3\}^{2n}$ defines (together with $S$) a valid Clifford operator. The set of $p$ for a given $S$ that correspond to a Clifford operator is given in Hostens \emph{et al.} \cite{hostens2005stabilizer} but it is of no consequence herein.}. It is easily shown that the symplectic matrix representing the composition of Clifford operator $U_1$ followed by Clifford operator $U_2$ is obtained by multiplying the symplectic matrices:
\begin{equation}
S(U_2U_1) = S(U_2)S(U_1).
\end{equation}
The expression for the phase vector for a composite gate is more complicated, and it is not needed herein (see Hostens \emph{et al.} \cite{hostens2005stabilizer} for the formula). The formula for parallel composition of Clifford operators in this representation is equally simple: because a gate can only transform the Pauli operators of the qubits on which it acts, the $S$ matrix for a gate acting on a subset of $n$ qubits is simply embedded into the identity on all other qubits. 

\subsection{The symplectic action of standard Clifford gates}
Our compilation algorithm is expressed in terms of the Hadamard ($H$), phase ($P$) and \textsc{cnot} gates:
\begin{align}
H \ket{x} &= \frac{1}{\sqrt{2}} (\ket{0} + (-1)^{x}\ket{1}),\\
P \ket{x} &= i^x\ket{x}, \\
\textsc{cnot}\, \ket{x,y} & = \ket{x,x+y},%\\
\end{align}
where $x,y=0,1$. A direct calculation can be used to show that the symplectic matrices corresponding to these gates are:
\begin{align}
S(H)  &= \begin{pmatrix} 0 & 1 \\ 1 & 0 \end{pmatrix},\\
 S(P)  &= \begin{pmatrix} 1 & 0 \\ 1 & 1 \end{pmatrix},\\
 S(\textsc{cnot})  &= \begin{pmatrix} 1 & 0 & 0 & 0 \\ 1 & 1 & 0 & 0 \\  0 & 0 & 1 & 1 \\  0 & 0 & 0 & 1  \end{pmatrix}.% \\
\end{align}
The corresponding phase vectors are the all-zeros vector, except for the phase gate for which $p(P) = (1,0)$. However, these phase vectors are only of secondary importance for our purposes. 

\begin{table}
\begin{footnotesize}
\begin{tabular}{ c | c | c    }
Gate &  Row (LHS) operation & Column (RHS) operation  \\
\hline\hline
$H_q$ & swaps rows $q$ and $q+n$  & swaps columns $q$ and $q+n$ \\
\hline
$P_q$ & adds row $q$ to $q + n$ & adds column $q+n$ to $q$ \\
\hline
$\textsc{cnot}_{q_1 \to q_2}$ & adds row $q_1$ to $q_2$  & adds column $q_2$ to $q_1$ \\
 &  adds row $q_2 + n$ to $q_1 + n$ & adds column $q_1 + n$ to $q_2 + n$ \\
\end{tabular}
\end{footnotesize}
\caption{The symplectic action of Hadamard, phase, and $\textsc{cnot}$ gates. This table shows the transformation that a Hadamard gate on qubit $q$ ($H_q$), a phase gate on qubit $q$ ($P_q$) and a \textsc{cnot} gate with control and target qubits $q_1$ and $q_2$, respectively, produces when applied to a $2n \times 2n$ symplectic matrix $S$ representing an $n$-qubit Clifford operator. The columns show the actions of the gates when applied to the left hand side (LHS, row action), or the right hand side (RHS, column action) of $U$. For example, when a Hadamard gate on a qubit $q$ is multiplied on the LHS (RHS) of a Clifford unitary $U$, the symplectic matrix for the composite unitary is given by swapping the $q$\textsuperscript{th} and $(q+n)$\textsuperscript{th} rows (columns) of $S(U)$.}
\label{table:action}
\end{table}

Our algorithm for compiling a Clifford operator $U$ is derived by left- and right-multiplying $U$ by circuits containing only $H$, $P$, or $\textsc{cnot}$ gates, until we have transformed $U$ into the identity operator. In the symplectic representation, multiplication of $U$ from the left-hand-side (LHS) or right-hand-side (RHS) by $H$, $P$, and $\textsc{cnot}$ gates applies the matrix transformations to $S(U)$ shown in Table~\ref{table:action}. Any \textsc{cnot} circuit $\c{C}$ implements a unitary that is represented by a symplectic matrix of the form \cite{aaronson2004improved}
\begin{equation}
S(\c{C})=\begin{pmatrix} M & 0 \\ 0 & M^{-T} \end{pmatrix}, \label{eq:cnot-circuit-notation}
\end{equation}
for an invertible matrix $M \in \mathbb{F}_2^{n\times n}$, where 
\begin{equation}
M^{-T} \equiv (M^{-1})^T,
\end{equation}
and inversion is with respect to $\mathbb{F}_2$. That is, $M^{-1}$ is the unique matrix in $\mathbb{F}_2^{n\times n}$ such that $MM^{-1} = I$, when such a matrix exists.
Moreover, every invertible matrix $M \in \mathbb{F}_2$ corresponds to some \textsc{cnot} circuit, which follows because \textsc{cnot} is universal for classical reversible logic. For convenience, we say that the \textsc{cnot} circuit implements $M$ if it satisfies Eq.~\eqref{eq:cnot-circuit-notation}, and if a \textsc{cnot} circuit is unspecified except that it is required to implement $M$ we denote it by $\c{C}_{M}$.

\subsection{A useful matrix property}
To derive our compiler we require the following property of matrices over $\mathbb{F}_2$:\cite{aaronson2004improved}

\vspace{0.2cm}
\noindent
 \emph{Property 1.}
For any symmetric matrix $A \in \mathbb{F}_2^{n \times n}$, there exists a diagonal matrix $B \in \mathbb{F}_2^{n \times n}$ such that $A + B = MM^T$ where $M \in \mathbb{F}_2^{n \times n}$ is invertible. 

\section{Clifford compilation}
Consider some arbitrary Clifford operator $U$ that we wish to compile, with $S \equiv S(U)$ and $p\equiv p(U)$ the symplectic representation of this Clifford operator. A circuit $\c{C}$ such that $S(\c{C})= S(U)$ implements a unitary that can differ from the desired Clifford operator only by an $n$-qubit Pauli operator. We therefore focus on finding a circuit $\c{C}$ with the property that $S(\c{C})=S(U)$. The formula for finding the Pauli layer to prefix or append to the circuit in order to obtain the correct Clifford operator is given in, e.g., Hostens \emph{et al.} \cite{hostens2005stabilizer}.

The derivation of our algorithm follows a similar format to that of Aaronson and Gottesman~\cite{aaronson2004improved}. First, we write $S$ as
\begin{equation} 
S = \begin{pmatrix} A & B \\ C & D \end{pmatrix} ,
\label{eq:s0}
\end{equation}
where $A$, $B$, $C$, and $D$ are $n \times n$ matrices over $\mathbb{F}_2$. Then we (1) use $\textsc{cnot}$, Hadamard and phase gate row operations and column operations on $S$ to map $S \to I$, and (2) reverse the order of these operations to map $I \to S$. We will act on $S$ from the left hand side (LHS) and the right hand side (RHS) with circuits containing either only $H$, $P$ or \textsc{CNOT} gates. These LHS and RHS operations implement row and column operations on $S$, respectively. We map $S \to I$ using the following 11 steps:

\begin{enumerate}
\item The matrices $A$, $B$, $C$ and $D$ in Eq.~\eqref{eq:s0} are not guaranteed to be invertible. However Lemma 2 of Aaronson and Gottesman~\cite{aaronson2004improved} implies that there exists a Hadamard circuit $\c{H}$ such that
\begin{equation}
S(\c{H}) \begin{pmatrix} A & B \\ C & D \end{pmatrix} = \begin{pmatrix} A_1 & Q \\ C_1 & D_1 \end{pmatrix},
\label{eq:s1}
 \end{equation}
for some $A_1$, $Q$, $C_1$, and $D_1$ with $Q$ invertible. 

\item As $Q$ is invertible there is a \textsc{cnot} circuit, $\c{C}_{Q^T}$, implementing its transpose. Multiplying Eq.~\eqref{eq:s1}, from the RHS, by the symplectic matrix for this circuit we obtain
\begin{equation}
\begin{pmatrix} A_1 & Q \\ C_1 & D_1 \end{pmatrix} S(\c{C}_{Q^T}) =  \begin{pmatrix} A_2 & I \\ C_2 & D_2 \end{pmatrix},
\label{eq:s2}
\end{equation}
for some $A_2$, $C_2$, and $D_2$. The RHS of Eq.~\eqref{eq:s2} is necessarily a symplectic matrix, and applying the defining property of a symplectic matrix [Eq.~\eqref{eq:symp}] we find that $D_2^T - D_2 = 0$. Therefore $D_2$ is symmetric.

\item Phase gates acting from the LHS (row operations) can add any diagonal matrix to $D_2$. Therefore, Property 1 implies that there exists a circuit of phase gates $\c{P}_1$ such that
\begin{equation}
S(\c{P}_1) \begin{pmatrix} A_2 & I \\ C_2 & D_2 \end{pmatrix}  = \begin{pmatrix} A_3 & I \\ C_3 & N^{-T}N^{-1} \end{pmatrix},
\label{eq:s3}
\end{equation}
for some $A_3$, $C_3$, and some invertible $N \in \mathbb{F}_2^{n \times n}$. 

\item  By apply a \textsc{cnot} circuit that implements $N^{-1}$ (i.e., $\c{C}_{N^{-1}}$) from the LHS, we obtain
\begin{equation}
 S(\c{C}_{N^{-1}})\begin{pmatrix} A_3 & I \\ C_3 & N^{-T}N^{-1} \end{pmatrix}  = \begin{pmatrix} A_4 & N^{-1} \\ C_4 & N^{-1} \end{pmatrix}.
\label{eq:s4}
\end{equation}

\item By applying a \textsc{cnot} circuit $\c{C}_{N^{-T}}$ from the RHS, we find that
\begin{equation}
\begin{pmatrix} A_4 & N^{-1} \\ C_4 & N^{-1} \end{pmatrix} S(\c{C}_{N^{-T}}) = \begin{pmatrix} A_5 & I \\ C_5 & I \end{pmatrix}.
\label{eq:s5}
\end{equation}

\item Applying the circuit $\c{P}_{\text{all}}$, consisting of a phase gate acting on every qubit, from the LHS, adds the upper matrices to the lower matrices. Therefore:
\begin{equation}
 S(\c{P}_{\text{all}})\begin{pmatrix} A_5 & I \\ C_5 & I \end{pmatrix} =\begin{pmatrix} A_6 & I \\ C_6 & 0 \end{pmatrix},
\label{eq:s6}
\end{equation}
where $A_6 = A_5$ and $C_6 = A_5 + C_5$. Applying the defining property of a symplectic matrix, we find that $C_6 = I$ and $A_6-A_6^T  = 0$, so $A_6$ is symmetric.

\item Next we swap the upper and lower matrices, by applying the circuit $\c{H}_{\text{all}}$, consisting of a Hadamard gate acting on every qubit, from the LHS:
\begin{equation}
 S(\c{H}_{\text{all}}) \begin{pmatrix} A_6 & I \\  I  & 0 \end{pmatrix} =\begin{pmatrix} I & 0 \\  A_6 & I \end{pmatrix}.
\label{eq:s7}
\end{equation} 

\item As $A_6$ is symmetric, Property 1 implies that there exists a circuit of phase gates $P_2$ such that
\begin{equation}
    S(\c{P}_2) \begin{pmatrix} I & 0 \\  A_6 & I \end{pmatrix} = \begin{pmatrix} I & 0 \\  M^{-T}M^{-1} & I \end{pmatrix},
    \label{eq:s8}
\end{equation}
for some invertible $M \in \mathbb{F}_2^{n \times n}$.
\item  Multiplying from the RHS by a \textsc{cnot} circuit $\c{C}_{M}$ that implements $M$, we obtain
\begin{equation}
\begin{pmatrix} I & 0 \\  M^{-T}M^{-1} & I \end{pmatrix} S(\c{C}_{M})  = \begin{pmatrix}   M & 0 \\ M^{-T} & M^{-T}\end{pmatrix}.
\label{eq:s9}
\end{equation}

\item Applied from the RHS, the $\c{P}_{\text{all}}$ circuit adds the RHS matrices to the LHS matrices. Therefore:
\begin{equation}
\begin{pmatrix}   M & 0 \\ M^{-T} & M^{-T}  \end{pmatrix} S(\c{P}_{\text{all}})  =\begin{pmatrix}  M & 0 \\ 0 & M^{-T}  \end{pmatrix}.
\label{eq:s10}
\end{equation}

\item The RHS of Eq.~\eqref{eq:s10} is a \textsc{cnot} circuit implenting $M$, and it is converted to the identity by multiplying by a \textsc{cnot} circuit $\c{C}_{M^{-1}}$  from the LHS (or RHS):
\begin{equation}
  S(\c{C}_{M^{-1} })\begin{pmatrix} M & 0\\ 0 & M^{-T}  \end{pmatrix} =\begin{pmatrix} I & 0\\ 0 & I \end{pmatrix}.
\label{eq:s11}
\end{equation}
\end{enumerate}

We now piece together the steps 1-11, shown in Eqs.~\eqref{eq:s1}-\eqref{eq:s11}, to obtain our complete algorithm. We use circuit composition notation whereby $\c{C}_1\mhyphen\c{C}_2$ denotes the circuit $\textsc{C}_1$ followed by the circuit $\c{C}_2$. In the above steps, there are three $\textsc{cnot}$ circuits that act sequentially from the RHS (see steps 2, 5, and 9), which can be combined into a single circuit. Specifically, $\c{C}_{M} \mhyphen \c{C}_{N^{-T}}  \mhyphen \c{C}_{Q^T} =  \c{C}_{L^{-1}}$ where
\begin{equation}
L = Q^{-T} N^{T} M^{-1}.
\end{equation}
Using this composite \textsc{cnot} circuit, we can write steps 1-11 mapping $S(U) \to I$ as the equality
\begin{multline}
S(\c{C}_{M^{-1}}) S( \c{P}_{2}) S( \c{H}_{\text{all}})  S(\c{P}_{\text{all}}) S(\c{C}_{N^{-1}} ) S(\c{P}_1) S( \c{H}) \times \\ S(U) S(\c{C}_{L^{-1}})S(\c{P}_{\text{all}}) = I,
\end{multline}
or, equivalently, as 
\begin{equation}
 S( \c{H} \mhyphen \c{P}_1 \mhyphen \c{C}_{N^{-1}} \mhyphen \c{P}_{\text{all}}  \mhyphen \c{H}_{\text{all}} \mhyphen \c{P}_2 \mhyphen \c{C}_{M^{-1}})  S(U) S(\c{P}_{\text{all}} \mhyphen \c{C}_{L^{-1}})  = I.
\end{equation}

Hadamard, phase and \textsc{cnot} gates are all equal to their inverses multiplied by some Pauli operator, and to invert a \textsc{cnot} implementing $K$ we simply run a \textsc{cnot} circuit implementing $K^{-1}$ (the reverse circuit is sufficient). Therefore, we can invert the above equation to obtain a compilation for any Clifford operator $U$:
\begin{equation}
 U   \sim \c{C}_{L} \mhyphen \c{P}_{\text{all}} \mhyphen  \c{C}_{M} \mhyphen \c{P}_2 \mhyphen \c{H}_{\text{all}}  \mhyphen \c{P}_{\text{all}} \mhyphen \c{C}_{N} \mhyphen \c{P}_1 \mhyphen \c{H},
\label{eq:clifford-compiler}
\end{equation}
where $U \sim \c{C}$ denotes that the circuit $\c{C}$ implements the unitary $U$ multiplied by some $n$-qubit Pauli operator.  This is our main result. The $L$, $N$ and $M$ matrices, and the $\c{P}_1$, $\c{P}_2$, and $\c{H}$ layers, can be derived from $U$ using steps 1-11 above (and they are not unique).

Bravyi and Maslov \cite{Bravyi2020-hg} present a decomposition of any Clifford operator $U$ into two $\textsc{cphase}$ circuits and one $\textsc{cnot}$ circuit, together with layers of single qubits gates, where 
\begin{equation}
\textsc{cphase}\, \ket{x,y}  = (-1)^{xy}\ket{x,y}.
\end{equation}
In particular, Bravyi and Maslov\cite{Bravyi2020-hg} show that any Clifford operator can be implemented by a circuit of the form $ \c{X} \mhyphen \c{Z} \mhyphen \c{P} \mhyphen \c{CX} \mhyphen \c{CZ} \mhyphen \c{H} \mhyphen \c{CZ} \mhyphen \c{H} \mhyphen \c{P}$, where $\c{CZ}$, $\c{CX}$, $\c{P}$, $\c{H}$, $\c{X}$, and $\c{Z}$ denote arbitrary circuits containing only \textsc{cphase}, \textsc{cnot}, $P$, $H$, $X$ and $Z$ gates, respectively. Circuits of only \textsc{cphase} gates implement a smaller class of functions than \textsc{cnot} circuits, and therefore a decomposition in which a $\c{CX}$ circuit is replaced with a $\c{CZ}$ circuit might result in a compilation containing fewer two-qubit gates on average. The decomposition of Eq.~\eqref{eq:clifford-compiler} (which contains three \textsc{cnot} circuits) can be transformed into a decomposition containing one \textsc{cnot} circuit and two $\textsc{cphase}$ circuits. To do so, we use two circuit transformation relations:\cite{maslov2017optimized}
\begin{align}
\c{CX} \mhyphen \c{H}_{\textrm{all}} &= \c{H}_{\textrm{all}} \mhyphen \c{CX}, \label{eq:com2}\\
\c{P} \mhyphen \c{CX} &\mapsto \c{CX} \mhyphen \c{CZ}  \mhyphen \c{P}, \label{eq:com1}
\end{align}
where the $\c{CX}$ and $\c{P}$ circuits that appear on each side of these relations are (in general) different, and where $\c{A} \mapsto \c{B}$ denotes that any circuit with the form $\c{A}$ can be mapped to a circuit with the form $\c{B}$ such that unitaries implemented by $\c{A}$ and $\c{B}$ differ only by an $n$-qubit Pauli operator. That is, Eq.~\eqref{eq:com1} denotes that any $\c{P} \mhyphen \c{CX}$ circuit can be rewritten in the form $\c{CX} \mhyphen \c{CZ} \mhyphen \c{P}$ (the converse is not true).\footnote{Note that Eq.~\eqref{eq:com1} can be derived by showing that $\c{P} \mhyphen \c{CX} \mapsto \c{CZ} \mhyphen \c{CX}  \mhyphen \c{P}$ and $\c{CZ} \mhyphen \c{CX} \mapsto \c{CX} \mhyphen \c{CZ}$. Both of these relations can be shown by checking that they hold for unit depth $\c{CX}$ and $\c{CZ}$ circuits, and then repeatedly applying that result.} By applying the circuit transformation relations of Eqs.~\eqref{eq:com2}-\eqref{eq:com1} to Eq.~\eqref{eq:clifford-compiler}, we obtain
\begin{align}
    \c{C}_{L} \mhyphen \c{P}_{\text{all}} \mhyphen  \c{C}_{M} & \mhyphen \c{P}_2 \mhyphen  \c{H}_{\text{all}}\mhyphen \c{P}_{\text{all}} \mhyphen \c{C}_{N} \mhyphen \c{P}_1 \mhyphen \c{H} \\ &\mapsto
    \c{C}_{L} \mhyphen \c{P}_{\text{all}} \mhyphen  \c{C}_{M} \mhyphen \c{P}_2  \mhyphen \c{H}_{\text{all}}  \mhyphen \c{CX} \mhyphen \c{CZ} \mhyphen \c{P} \mhyphen \c{H} 
    \\ &\mapsto
    \c{C}_{L} \mhyphen \c{P}_{\text{all}} \mhyphen  \c{C}_{M} \mhyphen \c{P}_2 \mhyphen  \c{CX}   \mhyphen \c{H}_{\text{all}} \mhyphen \c{CZ} \mhyphen \c{P} \mhyphen \c{H} 
    \\ &\mapsto
    \c{C}_{L} \mhyphen \c{P}_{\text{all}} \mhyphen  \c{CX} \mhyphen  \c{CZ} \mhyphen \c{P}   \mhyphen \c{H}_{\text{all}} \mhyphen \c{CZ} \mhyphen \c{P} \mhyphen \c{H} 
    \\ &\mapsto
    \c{CX} \mhyphen  \c{CZ} \mhyphen \c{P}   \mhyphen \c{H}_{\text{all}} \mhyphen \c{CZ} \mhyphen \c{P} \mhyphen \c{H}.
\end{align}
In conclusion,
\begin{equation}
 U   \sim \c{CX} \mhyphen \c{CZ} \mhyphen \c{P}  \mhyphen \c{H}_{\text{all}}  \mhyphen \c{CZ} \mhyphen \c{P} \mhyphen \c{H},
\label{eq:clifford-compiler-2}
\end{equation}
which is a compilation for any Clifford operator $U$ in terms of one $\textsc{cnot}$ circuit and two $\textsc{cphase}$ circuits. The exact form of each sub-circuit in Eq.~\eqref{eq:clifford-compiler-2} can be computed by applying the commutation relations in Eqs.~\eqref{eq:com1}-\eqref{eq:com2} to the fully-specified circuit in Eq.~\eqref{eq:clifford-compiler}.

\section{Stabilizer state compilation} 
The Clifford circuit compilation algorithm summarized by Eq.~\eqref{eq:clifford-compiler} can be used for efficient stabilizer state creation. In particular, it implies that any $n$-qubit stabilizer state can be created using a single \textsc{cnot} circuit and local gates. This is because, when acting on the $\ket{0}^n$ state, the first four subcircuits in Eq.~\eqref{eq:clifford-compiler} [i.e., the leftmost subcircuits] have no effect. Therefore, if we express an $n$-qubit stabilizer state $\ket{\psi}$ by $\ket{\psi} = U\ket{0}^n$ for some Clifford operator $U$, we can create that stabilizer state from $\ket{0}^n$ using the circuit
\begin{equation}
 \c{C}_{\psi} =   \c{H}_{\text{all}}  \mhyphen \c{P}_{\text{all}} \mhyphen \c{C}_{N} \mhyphen  \c{F} ,
\end{equation}
where $\c{F}$ is a circuit of one-qubit gates, consisting of $\c{P}_1 \mhyphen \c{H}$ followed by a layer of Pauli operator, and $N$ is the matrix specified in the algorithm for compiling $U$. Similarly, using Eq.~\eqref{eq:clifford-compiler-2}, we can create an arbitrary $n$-qubit stabilizer state $\ket{\psi}$ from $\ket{0}^n$ using the circuit
\begin{equation}
 \c{C}_{\psi} =   \c{H}_{\text{all}}   \mhyphen \c{CZ} \mhyphen  \c{F}' ,
\end{equation}
where $\c{F}'$ is a circuit of one-qubit gates. Note that single $\c{CX}$ or $\c{CZ}$ circuits for generating stabilizer states also immediately follow from Bravyi and Maslov's algorithm \cite{Bravyi2020-hg}. Furthermore, note that Aaronson and Gottesman \cite{aaronson2004improved} pointed out that any $n$-qubit stabilizer state can be generated from a single \textsc{cnot} circuit and local gates, but that work does not provide an explicit compilation algorithm. 

\begin{acknowledgments}
We thank Dmitri Maslov and Sergey Bravyi for useful comments on a draft of this manuscript, and for finding and helping to resolve a technical error in our derivation.
This research was funded, in part, by the Office of the Director of National Intelligence (ODNI), Intelligence Advanced Research Projects Activity (IARPA). All statements of fact, opinion, or conclusions contained herein are
those of the authors and should not be construed as representing the official views or policies of IARPA, the ODNI, or the U.S. Government. Sandia National Laboratories is a multi-mission laboratory managed and operated by National Technology \& Engineering Solutions of Sandia, LLC (NTESS), a wholly owned subsidiary of Honeywell International Inc., for the U.S. Department of Energy’s National Nuclear Security Administration (DOE/NNSA) under contract DE-NA0003525. This written work is authored by an employee of NTESS. The employee, not NTESS, owns the right, title and interest in and to the written work and is responsible for its contents. Any subjective views or opinions that might be expressed in the written work do not necessarily represent the views of the U.S. Government. The publisher acknowledges that the U.S. Government retains a non-exclusive, paid-up, irrevocable, world-wide license to publish or reproduce the published form of this written work or allow others to do so, for U.S. Government purposes. The DOE will provide public access to results of federally sponsored research in accordance with the DOE Public Access Plan.
\end{acknowledgments}

\bibliography{Bibliography}

\begin{thebibliography}{11}%
\makeatletter
\providecommand \@ifxundefined [1]{%
 \@ifx{#1\undefined}
}%
\providecommand \@ifnum [1]{%
 \ifnum #1\expandafter \@firstoftwo
 \else \expandafter \@secondoftwo
 \fi
}%
\providecommand \@ifx [1]{%
 \ifx #1\expandafter \@firstoftwo
 \else \expandafter \@secondoftwo
 \fi
}%
\providecommand \natexlab [1]{#1}%
\providecommand \enquote  [1]{``#1''}%
\providecommand \bibnamefont  [1]{#1}%
\providecommand \bibfnamefont [1]{#1}%
\providecommand \citenamefont [1]{#1}%
\providecommand \href@noop [0]{\@secondoftwo}%
\providecommand \href [0]{\begingroup \@sanitize@url \@href}%
\providecommand \@href[1]{\@@startlink{#1}\@@href}%
\providecommand \@@href[1]{\endgroup#1\@@endlink}%
\providecommand \@sanitize@url [0]{\catcode `\\12\catcode `\$12\catcode
  `\&12\catcode `\#12\catcode `\^12\catcode `\_12\catcode `\%12\relax}%
\providecommand \@@startlink[1]{}%
\providecommand \@@endlink[0]{}%
\providecommand \url  [0]{\begingroup\@sanitize@url \@url }%
\providecommand \@url [1]{\endgroup\@href {#1}{\urlprefix }}%
\providecommand \urlprefix  [0]{URL }%
\providecommand \Eprint [0]{\href }%
\providecommand \doibase [0]{http://dx.doi.org/}%
\providecommand \selectlanguage [0]{\@gobble}%
\providecommand \bibinfo  [0]{\@secondoftwo}%
\providecommand \bibfield  [0]{\@secondoftwo}%
\providecommand \translation [1]{[#1]}%
\providecommand \BibitemOpen [0]{}%
\providecommand \bibitemStop [0]{}%
\providecommand \bibitemNoStop [0]{.\EOS\space}%
\providecommand \EOS [0]{\spacefactor3000\relax}%
\providecommand \BibitemShut  [1]{\csname bibitem#1\endcsname}%
\let\auto@bib@innerbib\@empty
%</preamble>
\bibitem [{\citenamefont {Dehaene}\ and\ \citenamefont
  {De~Moor}(2003)}]{dehaene2003clifford}%
  \BibitemOpen
  \bibfield  {author} {\bibinfo {author} {\bibfnamefont {J.}~\bibnamefont
  {Dehaene}}\ and\ \bibinfo {author} {\bibfnamefont {B.}~\bibnamefont
  {De~Moor}},\ }\bibfield  {title} {\enquote {\bibinfo {title} {Clifford group,
  stabilizer states, and linear and quadratic operations over {GF(2)}},}\
  }\href {https://journals.aps.org/pra/abstract/10.1103/PhysRevA.68.042318}
  {\bibfield  {journal} {\bibinfo  {journal} {Phys. Rev. A}\ }\textbf {\bibinfo
  {volume} {68}},\ \bibinfo {pages} {042318} (\bibinfo {year}
  {2003})}\BibitemShut {NoStop}%
\bibitem [{\citenamefont {Hostens}, \citenamefont {Dehaene},\ and\
  \citenamefont {De~Moor}(2005)}]{hostens2005stabilizer}%
  \BibitemOpen
  \bibfield  {author} {\bibinfo {author} {\bibfnamefont {E.}~\bibnamefont
  {Hostens}}, \bibinfo {author} {\bibfnamefont {J.}~\bibnamefont {Dehaene}}, \
  and\ \bibinfo {author} {\bibfnamefont {B.}~\bibnamefont {De~Moor}},\
  }\bibfield  {title} {\enquote {\bibinfo {title} {Stabilizer states and
  {C}lifford operations for systems of arbitrary dimensions and modular
  arithmetic},}\ }\href
  {https://journals.aps.org/pra/abstract/10.1103/PhysRevA.71.042315} {\bibfield
   {journal} {\bibinfo  {journal} {Phys. Rev. A}\ }\textbf {\bibinfo {volume}
  {71}},\ \bibinfo {pages} {042315} (\bibinfo {year} {2005})}\BibitemShut
  {NoStop}%
\bibitem [{\citenamefont {Patel}, \citenamefont {Markov},\ and\ \citenamefont
  {Hayes}(2008)}]{patel2003efficient}%
  \BibitemOpen
  \bibfield  {author} {\bibinfo {author} {\bibfnamefont {K.~N.}\ \bibnamefont
  {Patel}}, \bibinfo {author} {\bibfnamefont {I.~L.}\ \bibnamefont {Markov}}, \
  and\ \bibinfo {author} {\bibfnamefont {J.~P.}\ \bibnamefont {Hayes}},\
  }\bibfield  {title} {\enquote {\bibinfo {title} {Efficient synthesis of
  linear reversible circuits},}\ }\href
  {https://arxiv.org/abs/quant-ph/0302002} {\bibfield  {journal} {\bibinfo
  {journal} {Quantum Inf. Comput.}\ }\textbf {\bibinfo {volume} {8}},\ \bibinfo
  {pages} {282--294} (\bibinfo {year} {2008})}\BibitemShut {NoStop}%
\bibitem [{\citenamefont {Aaronson}\ and\ \citenamefont
  {Gottesman}(2004)}]{aaronson2004improved}%
  \BibitemOpen
  \bibfield  {author} {\bibinfo {author} {\bibfnamefont {S.}~\bibnamefont
  {Aaronson}}\ and\ \bibinfo {author} {\bibfnamefont {D.}~\bibnamefont
  {Gottesman}},\ }\bibfield  {title} {\enquote {\bibinfo {title} {Improved
  simulation of stabilizer circuits},}\ }\href
  {https://journals.aps.org/pra/abstract/10.1103/PhysRevA.70.052328} {\bibfield
   {journal} {\bibinfo  {journal} {Phys. Rev. A}\ }\textbf {\bibinfo {volume}
  {70}},\ \bibinfo {pages} {052328} (\bibinfo {year} {2004})}\BibitemShut
  {NoStop}%
\bibitem [{\citenamefont {Maslov}\ and\ \citenamefont
  {Roetteler}(2018)}]{maslov2017optimized}%
  \BibitemOpen
  \bibfield  {author} {\bibinfo {author} {\bibfnamefont {D.}~\bibnamefont
  {Maslov}}\ and\ \bibinfo {author} {\bibfnamefont {M.}~\bibnamefont
  {Roetteler}},\ }\bibfield  {title} {\enquote {\bibinfo {title} {Shorter
  stabilizer circuits via bruhat decomposition and quantum circuit
  transformations},}\ }\href {https://ieeexplore.ieee.org/document/8335339/}
  {\bibfield  {journal} {\bibinfo  {journal} {IEEE Trans. Inf. Theory}\
  }\textbf {\bibinfo {volume} {64(7)}},\ \bibinfo {pages} {4729--4738}
  (\bibinfo {year} {2018})}\BibitemShut {NoStop}%
\bibitem [{\citenamefont {Amy}, \citenamefont {Bennett-Gibbs},\ and\
  \citenamefont {Ross}()}]{Amy2022-gt}%
  \BibitemOpen
  \bibfield  {author} {\bibinfo {author} {\bibfnamefont {M.}~\bibnamefont
  {Amy}}, \bibinfo {author} {\bibfnamefont {O.}~\bibnamefont {Bennett-Gibbs}},
  \ and\ \bibinfo {author} {\bibfnamefont {N.~J.}\ \bibnamefont {Ross}},\
  }\bibfield  {title} {\enquote {\bibinfo {title} {Symbolic synthesis of
  {C}lifford circuits and beyond},}\ }\href {http://arxiv.org/abs/2204.14205}
  {\ }\Eprint {http://arxiv.org/abs/2204.14205} {arXiv:2204.14205 [quant-ph]}
  \BibitemShut {NoStop}%
\bibitem [{\citenamefont {Bravyi}\ and\ \citenamefont
  {Maslov}(2021)}]{Bravyi2020-hg}%
  \BibitemOpen
  \bibfield  {author} {\bibinfo {author} {\bibfnamefont {S.}~\bibnamefont
  {Bravyi}}\ and\ \bibinfo {author} {\bibfnamefont {D.}~\bibnamefont
  {Maslov}},\ }\bibfield  {title} {\enquote {\bibinfo {title} {Hadamard-free
  circuits expose the structure of the {C}lifford group},}\ }\href
  {https://ieeexplore.ieee.org/document/9435351/} {\bibfield  {journal}
  {\bibinfo  {journal} {IEEE Trans. Inf. Theory}\ }\textbf {\bibinfo {volume}
  {67(7)}},\ \bibinfo {pages} {4546--4563} (\bibinfo {year}
  {2021})}\BibitemShut {NoStop}%
\bibitem [{\citenamefont {Bravyi}\ \emph {et~al.}(2021)\citenamefont {Bravyi},
  \citenamefont {Shaydulin}, \citenamefont {Hu},\ and\ \citenamefont
  {Maslov}}]{Bravyi2021-my}%
  \BibitemOpen
  \bibfield  {author} {\bibinfo {author} {\bibfnamefont {S.}~\bibnamefont
  {Bravyi}}, \bibinfo {author} {\bibfnamefont {R.}~\bibnamefont {Shaydulin}},
  \bibinfo {author} {\bibfnamefont {S.}~\bibnamefont {Hu}}, \ and\ \bibinfo
  {author} {\bibfnamefont {D.}~\bibnamefont {Maslov}},\ }\bibfield  {title}
  {\enquote {\bibinfo {title} {Clifford circuit optimization with templates and
  symbolic {P}auli gates},}\ }\href {\doibase 10.22331/q-2021-11-16-580}
  {\bibfield  {journal} {\bibinfo  {journal} {Quantum}\ }\textbf {\bibinfo
  {volume} {5}},\ \bibinfo {pages} {580} (\bibinfo {year} {2021})}\BibitemShut
  {NoStop}%
\bibitem [{\citenamefont {Nielsen}\ \emph {et~al.}(2020)\citenamefont
  {Nielsen}, \citenamefont {Rudinger}, \citenamefont {Proctor}, \citenamefont
  {Russo}, \citenamefont {Young},\ and\ \citenamefont
  {Blume-Kohout}}]{nielsen2020probing}%
  \BibitemOpen
  \bibfield  {author} {\bibinfo {author} {\bibfnamefont {E.}~\bibnamefont
  {Nielsen}}, \bibinfo {author} {\bibfnamefont {K.}~\bibnamefont {Rudinger}},
  \bibinfo {author} {\bibfnamefont {T.}~\bibnamefont {Proctor}}, \bibinfo
  {author} {\bibfnamefont {A.}~\bibnamefont {Russo}}, \bibinfo {author}
  {\bibfnamefont {K.}~\bibnamefont {Young}}, \ and\ \bibinfo {author}
  {\bibfnamefont {R.}~\bibnamefont {Blume-Kohout}},\ }\bibfield  {title}
  {\enquote {\bibinfo {title} {Probing quantum processor performance with
  py{GST}i},}\ }\href
  {https://iopscience.iop.org/article/10.1088/2058-9565/ab8aa4} {\bibfield
  {journal} {\bibinfo  {journal} {Quantum Sci. Technol.}\ }\textbf {\bibinfo
  {volume} {5}},\ \bibinfo {pages} {044002} (\bibinfo {year}
  {2020})}\BibitemShut {NoStop}%
\bibitem [{\citenamefont {Proctor}\ \emph {et~al.}(2019)\citenamefont
  {Proctor}, \citenamefont {Carignan-Dugas}, \citenamefont {Rudinger},
  \citenamefont {Nielsen}, \citenamefont {Blume-Kohout},\ and\ \citenamefont
  {Young}}]{proctor2018direct}%
  \BibitemOpen
  \bibfield  {author} {\bibinfo {author} {\bibfnamefont {T.~J.}\ \bibnamefont
  {Proctor}}, \bibinfo {author} {\bibfnamefont {A.}~\bibnamefont
  {Carignan-Dugas}}, \bibinfo {author} {\bibfnamefont {K.}~\bibnamefont
  {Rudinger}}, \bibinfo {author} {\bibfnamefont {E.}~\bibnamefont {Nielsen}},
  \bibinfo {author} {\bibfnamefont {R.}~\bibnamefont {Blume-Kohout}}, \ and\
  \bibinfo {author} {\bibfnamefont {K.}~\bibnamefont {Young}},\ }\bibfield
  {title} {\enquote {\bibinfo {title} {Direct randomized benchmarking for
  multiqubit devices},}\ }\href
  {https://journals.aps.org/prl/abstract/10.1103/PhysRevLett.123.030503}
  {\bibfield  {journal} {\bibinfo  {journal} {Phys. Rev. Lett.}\ }\textbf
  {\bibinfo {volume} {123}} (\bibinfo {year} {2019})}\BibitemShut {NoStop}%
\bibitem [{\citenamefont {Magesan}, \citenamefont {Gambetta},\ and\
  \citenamefont {Emerson}(2011)}]{magesan2011scalable}%
  \BibitemOpen
  \bibfield  {author} {\bibinfo {author} {\bibfnamefont {E.}~\bibnamefont
  {Magesan}}, \bibinfo {author} {\bibfnamefont {J.~M.}\ \bibnamefont
  {Gambetta}}, \ and\ \bibinfo {author} {\bibfnamefont {J.}~\bibnamefont
  {Emerson}},\ }\bibfield  {title} {\enquote {\bibinfo {title} {Scalable and
  robust randomized benchmarking of quantum processes},}\ }\href
  {https://journals.aps.org/prl/abstract/10.1103/PhysRevLett.106.180504}
  {\bibfield  {journal} {\bibinfo  {journal} {Phys. Rev. Lett.}\ }\textbf
  {\bibinfo {volume} {106}},\ \bibinfo {pages} {180504} (\bibinfo {year}
  {2011})}\BibitemShut {NoStop}%
\end{thebibliography}%

\end{document}